\documentclass[twocolumn,pra,superscriptaddress]{revtex4}
\usepackage{graphicx}
\usepackage{amssymb,amsmath}
\usepackage{bm}
\usepackage{dcolumn}
\usepackage{subfigure}
\usepackage{float}
\usepackage{url}
\usepackage{color}
\usepackage{txfonts}
\usepackage[driverfallback=dvipdfm]{hyperref}
\hypersetup{pdfpagemode=FullScreen,colorlinks=true,breaklinks,urlcolor=blue,linkcolor=blue,citecolor=blue}
\usepackage{natbib}

\begin{document}

\title{Quantum Chaos for the Unitary Fermi Gas from the Generalized Boltzmann Equations}

\author{Pengfei Zhang}
\affiliation{Kavli Institute for Theoretical Physics, University of California, Santa Barbara, CA 93106, USA}
\affiliation{Institute for Advanced Study, Tsinghua University, Beijing, 100084, China}

\date{\today }

\begin{abstract}

In this paper, we study the chaotic behavior of the unitary Fermi gas in both high and low temperature limits by calculating the Quantum Lyapunov exponent defined in terms of the out-of-time-order correlator. We take the method of generalized Boltzmann equations derived from the augmented Keldysh approach \cite{augKeldysh}. At high temperature, the system is described by weakly interacting fermions with two spin components and the Lyapunov exponent is found to be $\lambda_L=21\frac{n}{T^{1/2}}$. Here $n$ is the density of fermions for a single spin component. In the low temperature limit, the system is a superfluid and can be described by phonon modes. Using the effective action derived in \cite{Son}, we find $\lambda_L=9\times 10^3\left(\frac{T}{T_F}\right)^4T$ where $T_F$ is the Fermi energy. By comparing these to existing results of heat conductivity, we find that $D_E\ll v^2 /\lambda_L$ where $D_E$ is the energy diffusion constant and $v$ is some typical velocity. We argue that this is related to the conservation law for such systems with quasi-particles.
\end{abstract}

\maketitle
\section{Introduction}
In recently years, the out-of-time-order correlator (OTOC), which is proposed to diagnose the quantum chaos, has drawn a lot of attention in both gravity, condensed matter and quantum information community. An OTOC $F_{WV}(t)$ for operator $W$ and $V$ with proper regularization is defined as \cite{Kitaev1,bh1,Larkin}: 
\begin{align}
F_{WV}(t)=\text{tr}\left[\sqrt{\rho}W^\dagger(t)V^\dagger(0)\sqrt{\rho}W(t)V(0)\right]/\mathcal{Z_\beta}. \label{OTOC}
\end{align}
Here $\rho=\exp(-\beta H)$  and $\mathcal{Z}_\beta$ is the thermal partition function. Let's consider systems with some small parameters which, for example, could be $1/N$ for a model with $N$ local degree of freedoms. For such systems, at an intermediate time scale, $F_{WV}(t)$ is believed to have an exponential deviation behavior $F_{WV}(t)\sim c_0-\epsilon \exp(\lambda_L t)$. Here $c_0$ is some constant and $\epsilon$ is the small parameter. $\lambda_L$ is defined as the quantum Lyapunov exponent and can be related to the classical Lyapunov exponent under semi-classical approximation \cite{Larkin}. An time scale, Lyapunov time $\tau_L$, can be defined as $1/\lambda_L$. Remarkably, the quantum Lyapunov exponent has been proved to be upper bounded by $2\pi/\beta$ for any quantum mechanical systems \cite{prove} and is saturated by models with gravity duals \cite{bh1,bh2,bh3}, including celebrated SYK models \cite{Kitaev2,SYK4,SYK1}. 

In condensed matter physics, an important related question is the exact relation between the information scrambling and the thermalization of a closed system. Although intuitively the information scrambling describes the loss of memories for a closed system which implies local thermal equilibrium, there are also examples where the thermalization time $\tau_{\text{eq}}\gg \tau_L$ \cite{upper Hartnoll}. Lyapunov expoenents are also found to be closely related to transport behaviors where some bounds are proposed for general diffusion constants \cite{lower Hartnoll,lower Blake,upper Hartnoll,upper Lucas}. Moreover, it is found that the relation $v_B^2\tau_L\sim D_E$ holds for holographic models \cite{Blake} and SYK chains\cite{Chain} where $v_B$ is the speed of information spreading and $D_E$ is the energy diffusion constant. Whether similar relations hold for realistic models is an interesting question. 

To get some understanding of these problems, it is helpful to study the chaotic behavior of some realistic models, especially those with possible holographic description, and compare different time scales. In this paper, we do such analysis on the unitary Fermi gas, which is a strongly interacting realistic model widely studied theoretically and experimentally. The ratio between the shear viscosity $\eta$ and entropy density $s$ measured in experiments \cite{exp0,exp} is the closest to the holographic bound $1/4\pi$ \cite{DTSon}. Moreover, the non-relativistic conformal symmetry of the unitary Fermi gas has been found to be compatible with the isometry of some classical geometry \cite{geometry1,geometry2}. Some evaluation of transport coefficients based on possible bulk description has been performed in \cite{transunitary}. 

It is difficult to study the unitary Fermi gas for arbitrary temperature due to the absence of a small parameter. One possible choice is to introduce large-N factors to suppress the quantum fluctuation. However this will not lead to a controlled calculation if we set $N$ finite finally. In this work we will focus on the high temperature limit and the low temperature limit where controlled analysis exists. The system can then be described by either dilute interacting fermions at high temperature or phonons in low temperature limit\cite{Son}. We use the method of generalized Boltzmann equations derived from augmented Keldysh approach \cite{augKeldysh} to study OTOCs. It is an analogy of traditional Boltzmann equation for the evolution of distribution functions \cite{Kamenev}, which predicts the behavior of the normal ordered correlators. This method has been shown \cite{graphene} to directly related to the Bethe-Salpeter equation method \cite{SYK1,stanford,graphene,critical,SK Jian,ON,phonon,diffusion,Dicke} for models with well-defined quasi-particles. As explained latter, the advantage of this method is the existence of a shortcut to directly writing out the generalized Boltzmann equations without field theory derivations. Since the Boltzmann equations exists even for classical systems, it would be interesting to study the reduction of quantum chaos to classical chaos by this method in the future.

The plan of this paper is the following. In section \ref{sec2} we firstly give a brief review of the path integral in augmented Keldysh contour using the example of the microscopic model for the unitary fermi gas. Then, using this path integral formula, we derive the generalized Boltzmann equations, which is the counterpart of the traditional Boltzmann equations in the traditional Keldysh approach with single forward and backward evolution. In section \ref{III} we give a shortcut to the generalized Boltzmann equations without a field-theory calculation in augmented Keldysh approach. We discuss some properties of generalized Boltzmann equations and present the results for the Lyapunov exponent for high temperature in section \ref{IV}. In section \ref{V}, we study the quantum chaos at low temperature using the effective phonon description. Some remarks and outlooks can be found in section \ref{VI}.

\section{the Generalized Boltzmann equations for the unitary Femi gas}\label{sec2}

 \begin{figure}[t]
 	\center
 	\includegraphics[width=0.8\columnwidth]{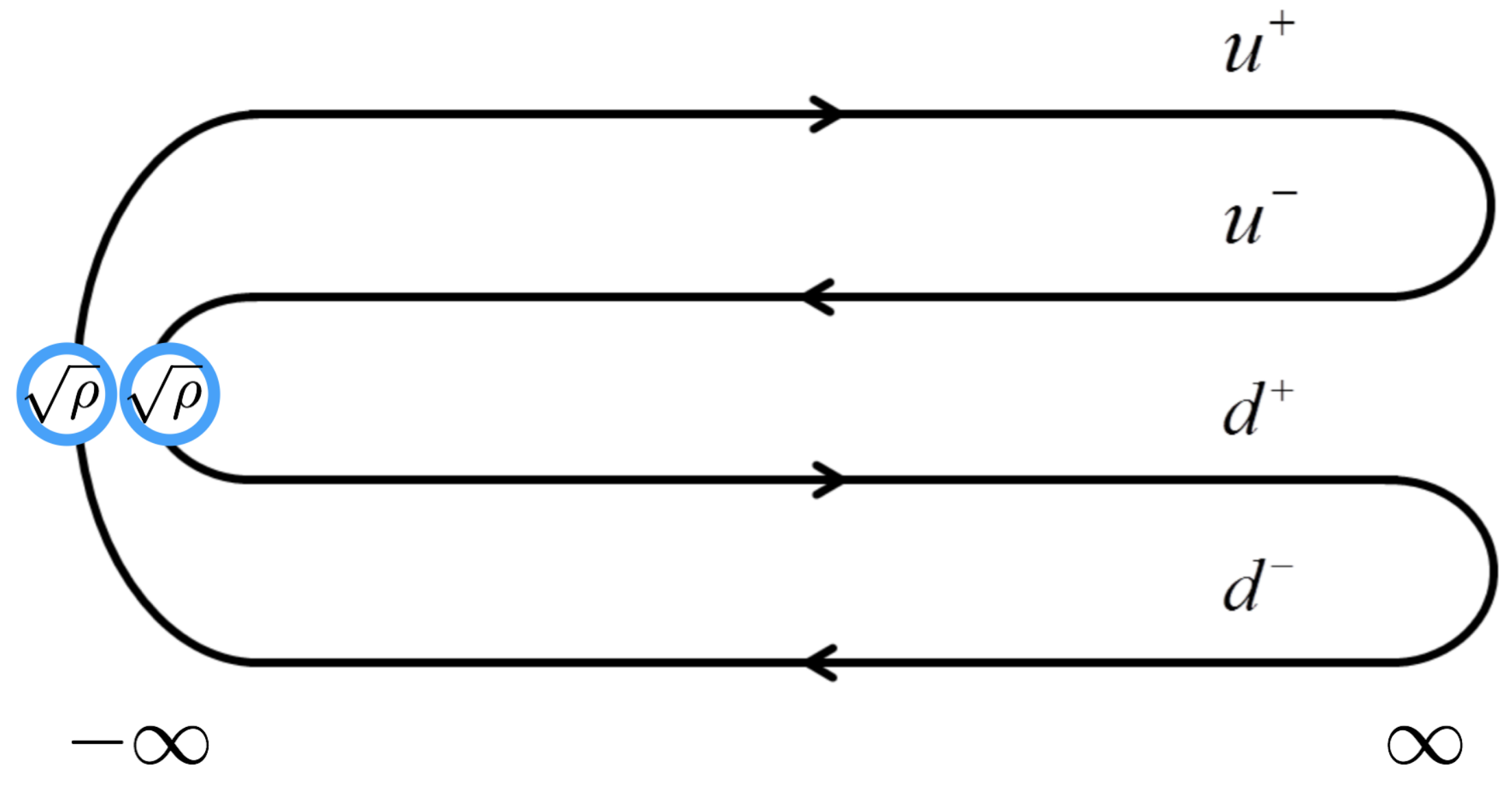}
 	\caption{A schematic of the augmented Keldysh contour used in this paper. }\label{fig:schematic}
 \end{figure}

In this section we give a brief review of the augmented Keldysh approach \cite{augKeldysh} and derive the generalized Boltzmann equations for contact interacting fermions. The relation between OTOC and th generalizeds Boltzmann equations is firstly proposed in \cite{augKeldysh}, and studied by adding a source term in \cite{graphene}. We also give some intuitive arguments in Appendix based on the evolution of thermal field doubled states \cite{TFD}.

To study OTOC \eqref{OTOC} which contains two forward/backward evolution and split thermal density matrix, one should double the time contour in the traditional Keldysh approach \cite{Kamenev} and insert a matrix element $\sqrt{\rho}$ between two copies ( different from the arrangement in \cite{Kamenev} ). A schematic of the time contour in shown in Fig. \ref{fig:schematic}, where we have labeled two copies by $u/d$ and forward/backward evolution by $+/-$. We then have four different fermion field $\psi_{\pm,u/d}$ for each spin specie $\uparrow$ or $\downarrow$. For the two component contact interacting fermions, the partition function reads:

\begin{align}
Z&=\int \mathcal{D}\psi\mathcal{D}\bar{\psi}\ \exp(i \int dt\ \mathcal{L}[\psi_{\pm,u/d}(x,t)]).\\
\mathcal{L}&=\int d\mathbf{x}\  \left(i\bar{\psi}(\hat{G^0})^{-1}\psi-g_s\sum_{m=u/d}\sum_{\xi=\pm}\xi\bar{\psi}_{\xi,m,\uparrow}\bar{\psi}_{\xi,m,\downarrow}\psi_{\xi,m,\downarrow}\psi_{\xi,m,\uparrow}\right).
\end{align}
Here we have omitted all indexes for fermion field in the $G^0$ term. The interaction term is diagonal in different contours with different sign for the forward and the backward evolution. The quadratic term leads to mixing between different contours because of the non-zero matrix element at $t=\pm \infty$. This similar to the traditional Keldysh approach case \cite{Kamenev}.

The determination of the bare Green's function $G^0_{\xi\eta,mm'}(\mathbf{x},t)=-i\left<\psi_{\xi,m}(\mathbf{x},t)\bar{\psi}_{\eta,m'}(\mathbf{0},0)\right>|_{g=0}$ (for each spin, for simplicity we just drop the spin index here.) can be largely simplified by realizing the unitarity of the time evolution operator which means we could contract forward and backward evolutions for redundant contours. As a result, we know that $G^0_{uu}=G^0_{dd}$ should be the same as the Green's function in the traditional Keldysh approach. After an Keldysh rotation defined by 
\begin{align}
\psi_{1,m}=\frac{1}{\sqrt{2}}(\psi_{+,m}+\psi_{-,m})\ \ \psi_{2,m}=\frac{1}{\sqrt{2}}(\psi_{+,m}-\psi_{-,m}),\\
\bar\psi_{1,m}=\frac{1}{\sqrt{2}}(\bar\psi_{+,m}-\bar\psi_{-,m})\ \ \bar\psi_{2,m}=\frac{1}{\sqrt{2}}(\bar\psi_{+,m}+\bar\psi_{-,m}),
\end{align}
we have $$G^0_{uu}=G^0_{dd}=\begin{pmatrix}
G^0_R&G^0_{K}\\
0&G^0_A
\end{pmatrix}.$$
The $G^0_R\ (G^0_A)$ is the bare retarded (advanced) Green's function for non-relativistic fermions given by $$G^0_R(\mathbf{p},\omega)=\left(G^0_A(\mathbf{p},\omega)\right)^*=\frac{1}{\omega-\mathbf{p}^2/2+i\epsilon}.$$ We have set the mass of fermions to be 1. For $G^0_K$ components, at thermal equilibrium we have a relation called Fluctuation-Dissipation Theorem (FDT): 
\begin{align}
G^0_{K}(\mathbf{p},\omega)=(G_R(\mathbf{p},\omega)-G_A(\mathbf{p},\omega))(1-2n_F(\omega,\mu)), \label{uu}
\end{align}
where $n_F(\omega,\mu)$ is the Fermi-Dirac distribution.

Now consider $G^0_{ud}$, it also has a simple structure in $1/2$ basis. The unitarity now implies only $G^0_{12,ud}\equiv G^0_{K,ud}$ is non-zero in the $2\times 2$ matrix of $G^0_{ud}$. It is straightforward to determine a generalized version of FDT:
\begin{align}
G^0_{K,ud}=-\frac{G_R(\mathbf{p},\omega)-G_A(\mathbf{p},\omega)}{\cosh(\beta\frac{\omega-\mu}{2})},\label{ud}
\end{align}
in thermal equilibrium by going back to the operator representation. Similarly, we have $G^0_{12,du}\equiv G^0_{K,du}=-G^0_{K,ud}$ for equilibrium system. Inversing the bare Green's function shows that all off-diagonal terms in $(\hat{G^0})^{-1}$ in $1/2$ basis are infinitely small and thus can be neglected when we add the self energy contribution. As a result, in real space and time, we could write $(\hat{G^0})^{-1}_{ab,m,m'}=\hat L_0\delta_{ab}\delta_{mm'}$ with $\hat L_0=i\partial_t+\nabla^2/2$. Here we have $a,b=1,2.$

To derive the generalized Boltzmann equation, let's consider our system with finite interaction strength $g$ is perturbed away from equilibrium. As a result the FDT does not hold and the system has no longer translational invariance. We write out the Schwinger-Dyson equation in real time:
\begin{align}
\left((\hat{G^0})^{-1}-\Sigma\right)\circ G=I.\label{SD}
\end{align}
Here we should keep in mind that the Green's function and self energy are all matrices of space, time, spin, $1/2$ and $u/d$. Since the unitarity still holds, the Greens function and the self energy still have the specific causal structure in $1/2$ index:
\begin{align}
G=\begin{pmatrix}
G_R&G_{K}\\
0&G_A
\end{pmatrix},\ \ \ \ \ \ \ \ 
\Sigma=\begin{pmatrix}
\Sigma_R&\Sigma_{K}\\
0&\Sigma_A
\end{pmatrix}.
\end{align}
and only $G_K$ and $\Sigma_{K}$ have non-vanishing matrix element between $u$ and $d$ contour. One could show that after defining $G_K=G_R\circ F-F\circ G_A$, which is motivated by FDT, Eq. \eqref{SD} leads to:
\begin{align}
L_0\circ F-F\circ L_0=(\Sigma_R\circ F-F \circ \Sigma_A)-\Sigma_K.
\end{align}
This is the same form as the results in the traditional Keldysh approach \cite{Kamenev}, although now each operator is a matrix with $u/d$ indexes. We call $F$ the distribution matrix.
\begin{figure}[t]
 	\center
 	\includegraphics[width=1\columnwidth]{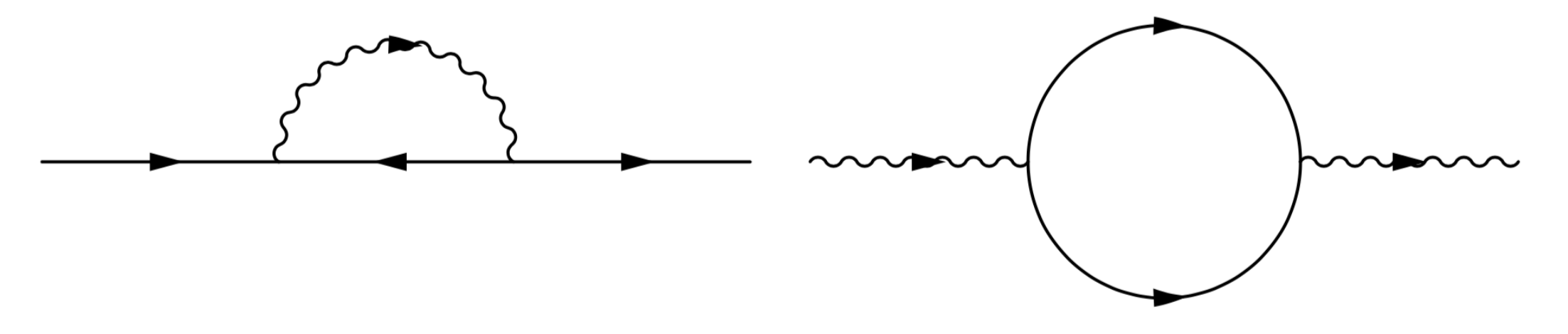}
 	\caption{Diagrams for the self energy of fermions and pairs that dominates in high temperature limit. Solid lines are propagators for fermions and wavy lines represents pairs.}\label{fig:Diagrams}
 \end{figure}

Up to now the derivation is exact. To proceed we need to take a semi-classical approximation by assuming a slow variation in both space and time. Mathematically, this calls for the Wigner transformation defined as:
\begin{align}
A(\mathbf{x},\mathbf{p})&=\mathcal{W}[A(\mathbf{x_1},\mathbf{x_2})]\notag\\&=\int \frac{d\mathbf{x_r}}{(2\pi)^3}A(\mathbf{x}+\frac{\mathbf{x_r}}{2},\mathbf{x}-\frac{\mathbf{x_r}}{2})\exp(i\mathbf{p}\cdot\mathbf{x_r}),
\end{align}
which separates center-of-mass coordinate $\mathbf{x}$ and semi-classical momentum $\mathbf{p}$ and simply reduces to Fourier transformation for systems with translational symmetry. Similar definition works for time and frequency space. To the leading order in fluctuation $(\nabla_x\cdot \nabla_p)$, we have:
\begin{align}
\mathcal{W}[A\circ B]\approx A B+\frac{i}{2}(\nabla_xA\cdot\nabla_pB-\nabla_pA\cdot\nabla_xB)+...
\end{align}
This expansion gives the Generalized Boltzmann equation for distribution matrix $F$ to the leading order:
\begin{align}
(\partial_t+\mathbf{p}\cdot\nabla_x)F(\mathbf{p})=i\Sigma_K-i(\Sigma_R F-F\Sigma_A)\equiv \text{St}[\mathbf{p},F]. \label{SDE}
\end{align}
Here we have defined $\text{St}[F]$ whose diagnal elements in $u/d$ space is proportional to collision integral \cite{Kinetic} in traditional Boltzmann equation.  We have also set the frequency in $F$ on-shell, which works to the leading order because $F$ always appears together with the spectral function. 

We now want to work out the explicit form of $\text{St}[F]$ for the unitary Fermi gas in high temperature limit. We firstly perform a Hubbard-Stratonovich transformation which introduces pair fields. The interaction part of the action is then given by:
\begin{align}
S_{\text{int}}&=-\int d\mathbf{x} dt \ \left(g_s\sum_{m,\xi}\xi\bar{\psi}_{\xi,m,\uparrow}\bar{\psi}_{\xi,m,\downarrow}\psi_{\xi,m,\downarrow}\psi_{\xi,m,\uparrow}\right)\notag\\
&=-\int d\mathbf{x} dt \ \sum_{m,\xi}\xi \left(\bar{\Delta}_{\xi,m}\psi_{\xi,m,\downarrow}\psi_{\xi,m,\uparrow}+\text{H.C.}-\frac{\bar{\Delta}_{\xi,m}\Delta_{\xi,m}}{g_s}\right)
\end{align}
Defining the classical and quantum components for pair fields:
\begin{align}
\Delta_{+,m}=\Delta_{\text{cl},m}+\Delta_{\text{q},m},\ \ \ \ \ \Delta_{-,m}=\Delta_{\text{cl},m}-\Delta_{\text{q},m},
\end{align}
the coupling between pairs and fermions becomes:
\begin{align}
-\int d\mathbf{x} dt\ \sum_{m,a}(\Delta_{a,m}(\psi_{m,\downarrow}^T\gamma^a\psi_{m,\uparrow})+\text{H.C.})
\end{align}
Here $a=\text{cl}$ or q. We have arranged annihilation operator of fermions into vectors in 1/2 space and defined $\gamma^{\text{cl}}=\sigma_1$ and $\gamma^{\text{q}}=\sigma_0$. We define the Green's function for bosons $g_{ab,mm'}(\mathbf{x},t)=-2i\left<\Delta_{a,m}(\mathbf{x},t)\bar\Delta_{b,m'}(\mathbf{0},0)\right>$ with the bare value $g^0_{ab,mm'}=g_s\delta_{mm'}\sigma_x$. The structure of the Green's function for bosonic fields is discussed in more details in \cite{Kamenev}, where one defines $g_{\text{cl}\text{cl}}=g_K$, $g_{\text{cl}\text{q}}=g_R$, $g_{\text{q}\text{cl}}=g_A$ and $g_{\text{q}\text{q}}=0$.

We consider the self-energy diagrams shown in Fig. \ref{fig:Diagrams}, which is the dominate contribution in high temperature limit for dilute gases \cite{Viral}. Alternatively, one could introduce large-$N$ indexes to suppress the fluctuation \cite{QPT}. The self energy matrix $\Sigma_{mm',\sigma}$ in with implicit 1/2 indexes is given by:
\begin{align}
-i&\Sigma_{mm',\sigma}(p)\notag\\&=-\int\sum_{a,b}\frac{d^4k}{(2\pi)^4}(-i\gamma^a)(iG^0_{m'm,-\sigma}(k-p))^T(-i\gamma^b)\frac{ig_{ab,mm'}(k)}{2}.\label{self1}
\end{align}
Here we define the self energy for pairs by $g^{-1}=(g_s^{-1}\sigma_x+\Pi)$, one could show that:
\begin{align}
g_K=-g_R\Pi_Kg_A, \ \ \ \ \ g_{R/A}=g_R(g_s^{-1}+\Pi_{A/R})g_{A},
\end{align}
with the self energy $\Pi_{mm',\sigma}$ given by:
\begin{align}
\Pi_{ab,mm'}(k)=-\frac{i}{2}\int\frac{d^4k'}{(2\pi)^4}\text{tr}\left[(i\gamma^a)iG^0_{mm',\uparrow}(k-k')(i\gamma^b)iG^0_{mm',\downarrow}(k')\right], \label{self2}
\end{align}
where $\Pi_{\text{clcl}}=0$, $\Pi_{\text{clq}}=\Pi_A$, $\Pi_{\text{qcl}}=\Pi_R$ and $\Pi_{\text{qq}}=\Pi_K$. Expanding in terms of  $z=\exp(\mu/T)$, which is valid in high temperature, we can approximate $g_{R,mm'}(k_0,\mathbf{k})=\frac{1}{1/4\pi a_s-i\sqrt{k_0-k^2/4}/4\pi}\delta_{mm'}$ which is the two body scattering matrix in vacuum by using the renormalization relation $$\frac{1}{g_s}=\frac{1}{4\pi a_s}+\int \frac{d^3k}{(2\pi)^3}\frac{1}{k^2},$$ which relates $g_s$ to the physical scattering length $a_s$. Straightforward derivations based on Eq. \eqref{self1} and \eqref{self2} lead to the final answer for $\text{St}[F]$. For $m=m'$, we have the scattering term in the traditional Boltzmann equation
\begin{align}
\text{St}_{mm}(\mathbf{p})=&\frac{1}{4}\int\frac{d\mathbf{p_1}d\mathbf{p_2}d\mathbf{p_3}}{(2\pi)^9}\mathcal{T}(\mathbf{p},\mathbf{p_i})(-\mathcal{L}_m(\mathbf{p_i}) F_{mm}(\mathbf{p})+F_{mm}(\mathbf{p_3})\notag\\&+F_{mm}(\mathbf{p_2})-F_{mm}(\mathbf{p_1})-F_{mm}(\mathbf{p_1})F_{mm}(\mathbf{p_2})F_{mm}(\mathbf{p_3})) \label{ori}
\end{align}
with 
\begin{align}
\mathcal{T}=\frac{(4\pi a_s)^2}{1+|\mathbf{p-p_1}|^2a_s^2/4}(2\pi)^4\delta^{(4)}(p+p_1-p_2-p_3，)
\end{align}
and
\begin{align}
\mathcal{L}_m=&F_{mm}(\mathbf{p_2})F_{mm}(\mathbf{p_3})-F_{mm}(\mathbf{p_1})F_{mm}(\mathbf{p_3})\notag\\&-F_{mm}(\mathbf{p_1})F_{mm}(\mathbf{p_2})+1
\end{align}
For simplicity we drop all time arguments for $F$. And for $m\neq m'$, the result for generalized Boltzmann equations is:
\begin{align}
\text{St}_{mm'}(\mathbf{p})=&\frac{1}{4}\int\frac{d\mathbf{p_1}d\mathbf{p_2}d\mathbf{p_3}}{(2\pi)^9}\mathcal{T}(\mathbf{p},\mathbf{p_i})(-(\mathcal{L}_m+\mathcal{L}_{m'})(\mathbf{p_i}) F_{mm'}(\mathbf{p})/2\notag\\&-F_{m'm}(\mathbf{p_1})F_{mm'}(\mathbf{p_2})F_{mm'}(\mathbf{p_3})).\label{genBol}
\end{align}
Here I have kept the spin index implicit, which can be put back by considering the scattering process. In this paper, we will only focus on a spin symmetric perturbation and $F$ is then spin-independent. These collision integrals are similar to the results for Fermi liquid in \cite{Kamenev}, although in which there are some typos. One could verify that $\text{St}[F]$ vanishes the for the equilibrium solution given bsy Eq. \eqref{uu} and \eqref{ud}.

\begin{figure}[t]
 	\center
 	\includegraphics[width=0.6\columnwidth]{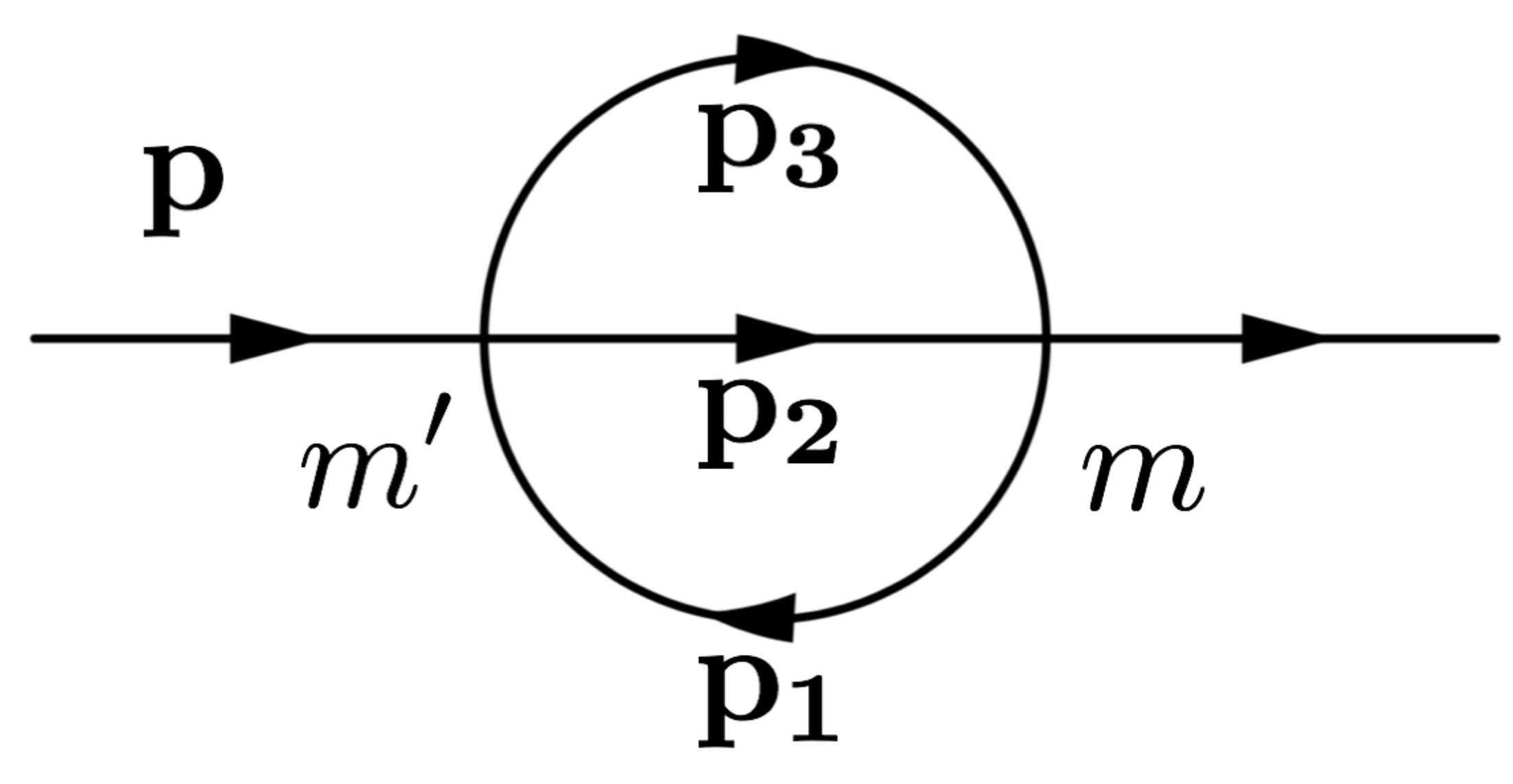}
 	\caption{The lowest order diagram for self eneregy with explicit momentum and $u/d$ label.}\label{fig:Diagrams2}
 \end{figure}

\section{A shortcut}\label{III}
One advantage of Boltzmann equation is that we do not need to repeat the derivation in Keldysh formalism every time because of its direct physical interpretation from Fermi's golden rule. What we need to know is only the transition rate which is given by the $T-$matrix. Similarly, here we want to present a short cut to the generalized Boltzmann equations, again based the same knowledge, to avoid the complicated derivations. 

Let's consider how terms appear in the collision integrals. Each term in the bracket can be traced back to Eq. \eqref{SDE}: The term proportional to $F(\mathbf{p})$ comes from $(\Sigma_R F-F\Sigma_A)$, which is more or less the same for either $uu/dd$ terms or $ud/du$ components since the contribution is always from the diagonal components of self-energy. Else terms are from $\Sigma_K$, which contain contributions from $G^0_R$ or $G^0_A$ for $m=m'$ while only the contribution from $G^0_K$ exists for $m\neq m'$. As a result, for $m\neq m'$ there is only one term $\sim F^3$. The label of $m$ and $m'$ in this term can be written out directly by considering the leading order scattering diagram shown in Fig. \ref{fig:Diagrams2} where we use the vertex before Hubbard-Stratonovich transformation with four-fermion interaction.

Based on these analysis, the short cut to the generalized Boltzmann equations can be summarized as follows:

1. Write out traditional Boltzmann equation in terms of distribution function $f(\mathbf{p},t)$ based on the transition rate, which counts the number of particles in certain momentum.

2. Define the variable $F(\mathbf{p},t)=1\mp2f(\mathbf{p},t)$ for fermions / bosons, translate the Boltzmann equation of $f(\mathbf{p},t)$ to the equation of $F(\mathbf{p},t)$. (This gives Eq. \eqref{ori}, where the factor of $1/4$ is because of the factor of 2 in the definition of $F$ above.)

3. Separate out the term proportional to $F(\mathbf p)$ in traditional Boltzmann equation as $\mathcal{L}[F] F(\mathbf p)$. Write a term $(\mathcal{L}[F_u]+\mathcal{L}[F_d]) F_{mm'}(\mathbf p)/2$ in the generalized Boltzmann equation.

4. For the remaining terms in traditional Boltzmann equation, only keep the term with largest number of $F$ and add labels of $mm'$ or $m'm$ to them as discussed previous by considering the leading order diagram.

One could straightforwardly show these rules lead to exactly the same equation as the one in Eq. \eqref{genBol}. Also one could verify that it is true for other models studied in \cite{Kamenev}. As an example, we will use them when analyzing the low-temperature chaotic behavior of the unitary Fermi gas.

\section{Quantum chaos in the high temperature limit}\label{IV}

Before proceeding to solve the generalized Boltzmann equation for the unitary Femi gas, we analyze some property of generalized Boltzmann equations using the example of interacting fermions. The evolution of diagonal terms in the distribution matrix $F_{uu}$ and $F_{dd}$ only depends on themselves. The existence of $H-$theorem \cite{Kinetic} guarantees their relaxation to the thermal equilibrium. For small deviation, one could linearize the Boltzmann equations. We could defined the deviation from thermal equilibrium by $$F_{mm'}(\mathbf{p})=F_{mm'}^0(\mathbf{p})+\delta F_{mm'}(\mathbf{p}).$$ One class of modes has both non-vanishing $\delta F_{uu}$ and $\delta F_{ud}$. The relaxation of them mean $\delta F_{uu,\alpha}(t,\mathbf{p})=\delta F_{dd,\alpha}(t,\mathbf{p})=-2\delta F_\alpha(t,\mathbf{p})=-2\delta_\alpha(\mathbf{p}) \exp(- \lambda_\alpha t)$ with $\lambda_\alpha >$0. At the same time, one could verify that the solution for $\delta F_{ud}$ and $\delta F_{du}$ is given by:
\begin{align}
&\delta F_{ud,\alpha}(t,\mathbf{p})=-2\delta F_\alpha(t,\mathbf{p})\exp(\beta\frac{\epsilon_p-\mu}{2}),\ \ \notag\\&\delta F_{du,\alpha}(t,\mathbf{p})=-2\delta F_\alpha(t,\mathbf{p})\exp(-\beta\frac{\epsilon_p-\mu}{2}).
\end{align}
Here $\alpha$ labels different solutions in this class. As a result all components of distribution matrices relax to equilibrium value. Another possible class of solutions satisfies $\delta F_{uu,\alpha}(t,\mathbf{p})=\delta F_{dd,\alpha}(t,\mathbf{p})=0$, which means the diagonal part is always in equilibrium. However, the off diagonal part is non trivial and given by:
\begin{align}
&\delta F_{ud,\alpha'}(t,\mathbf{p})=-2\delta_{ud,\alpha'}(\mathbf{p})\exp(\beta\frac{\epsilon_p-\mu}{2}) \exp(- \lambda_{\alpha'} t),\ \ \notag\\&\delta F_{du,\alpha'}(t,\mathbf{p})=-2\delta_{du,\alpha'}(\mathbf{p})\exp(-\beta\frac{\epsilon_p-\mu}{2}) \exp(- \lambda_{\alpha'} t). \label{class2}
\end{align}
However now $\lambda_{\alpha'}$ may becomes negative. For a general initial condition, the solution should be a superposition of all these eigenmodes. If the perturbation in $F_{ud}$ and $F_{du}$ is larger then that of $F_{uu}$ and $F_{dd}$, such that the coefficients of terms in Eq. \eqref{class2} are positive \footnote{This depend on short time behavior of the perturbation. In special set-up one could straightforwardly show this is true for free fermion systems.}, we expect an exponential deviation from thermal equilibrium for the off-diagonal components of the distribution matrix and the exponent, which is given by the negative $\lambda_{\alpha'}$ with the largest absolute value, gives the Lyapunov exponent $\lambda_L=\text{max}_{\lambda_{\alpha'}<0}|\lambda_{\alpha'}|$.  

With these understanding, we proceed to solve the generalized Boltzmann equations. Since we are interested in the Lyapunov exponent, we just set the diagonal components to be thermal equilibrium $F_{uu}^0(\mathbf{p})=F_{dd}^0(\mathbf{p})=(1-2n_F(p^2/2,\mu))\equiv F^0(\mathbf{p})$ \cite{augKeldysh,graphene}. We assume a spin-independent and spacial homogeneous initial condition. Linearizing the equation, we get:
\begin{align}
\text{St}_{mm'}&(\mathbf{p})=\frac{1}{4}\int\frac{d\mathbf{p_1}d\mathbf{p_2}d\mathbf{p_3}}{(2\pi)^9}\mathcal{T}(\mathbf{p},\mathbf{p_i})(-\mathcal{L}^0(\mathbf{p_i}) \delta F_{mm'}(\mathbf{p})\notag\\&-F^0_{m'm}(\mathbf{p_1})F^0_{mm'}(\mathbf{p_2})\delta F_{mm'}(\mathbf{p_3})-F^0_{m'm}(\mathbf{p_1})\delta F_{mm'}(\mathbf{p_2})\notag\\&F^0_{mm'}(\mathbf{p_3})-\delta F_{m'm}(\mathbf{p_1})F^0_{mm'}(\mathbf{p_2})F^0_{mm'}(\mathbf{p_3})),  \label{linear2}
\end{align}
with
\begin{align}
\mathcal{L}^0=&F^0(\mathbf{p_2})F^0(\mathbf{p_3})-F^0(\mathbf{p_1})F^0(\mathbf{p_3})-F^0(\mathbf{p_1})F^0(\mathbf{p_2})+1,
\end{align}
and $F_{ud}^0(\mathbf{p})=-F_{du}^0(\mathbf{p})=-\frac{1}{\cosh(\beta(p^2/2-\mu)/2)}.$  In an equivalent Bethe-Salpeter calculation, the first term in Eq. \eqref{linear2} corresponds to the self energy of Green's functions while other terms correspond to a convolution with kernels with one or two rungs.

Since the calculation is controlled in high temperature limit, we keep all terms to the leading order of $z$ in Eq. \eqref{linear2} and find:
\begin{align}
\text{St}_{mm'}(\mathbf{p})=&\int\frac{d\mathbf{p_1}d\mathbf{p_2}d\mathbf{p_3}}{(2\pi)^9}\frac{\mathcal{T}(\mathbf{p},\mathbf{p_i})}{4}(-4\exp(\frac{\mu-p_1^2/2}{T}) \delta F_{mm'}(\mathbf{p})\notag\\&-4\exp(\frac{\mu-p_2^2/4-p_3^2/4}{T}) \delta F_{mm'}(\mathbf{p_1}) \notag\\& +8\exp(\frac{\mu-p_1^2/4-p_3^2/4}{T}) \delta F_{mm'}(\mathbf{p_2})).  \label{linear2}
\end{align}

\begin{figure}[t]
 	\center
 	\includegraphics[width=0.9\columnwidth]{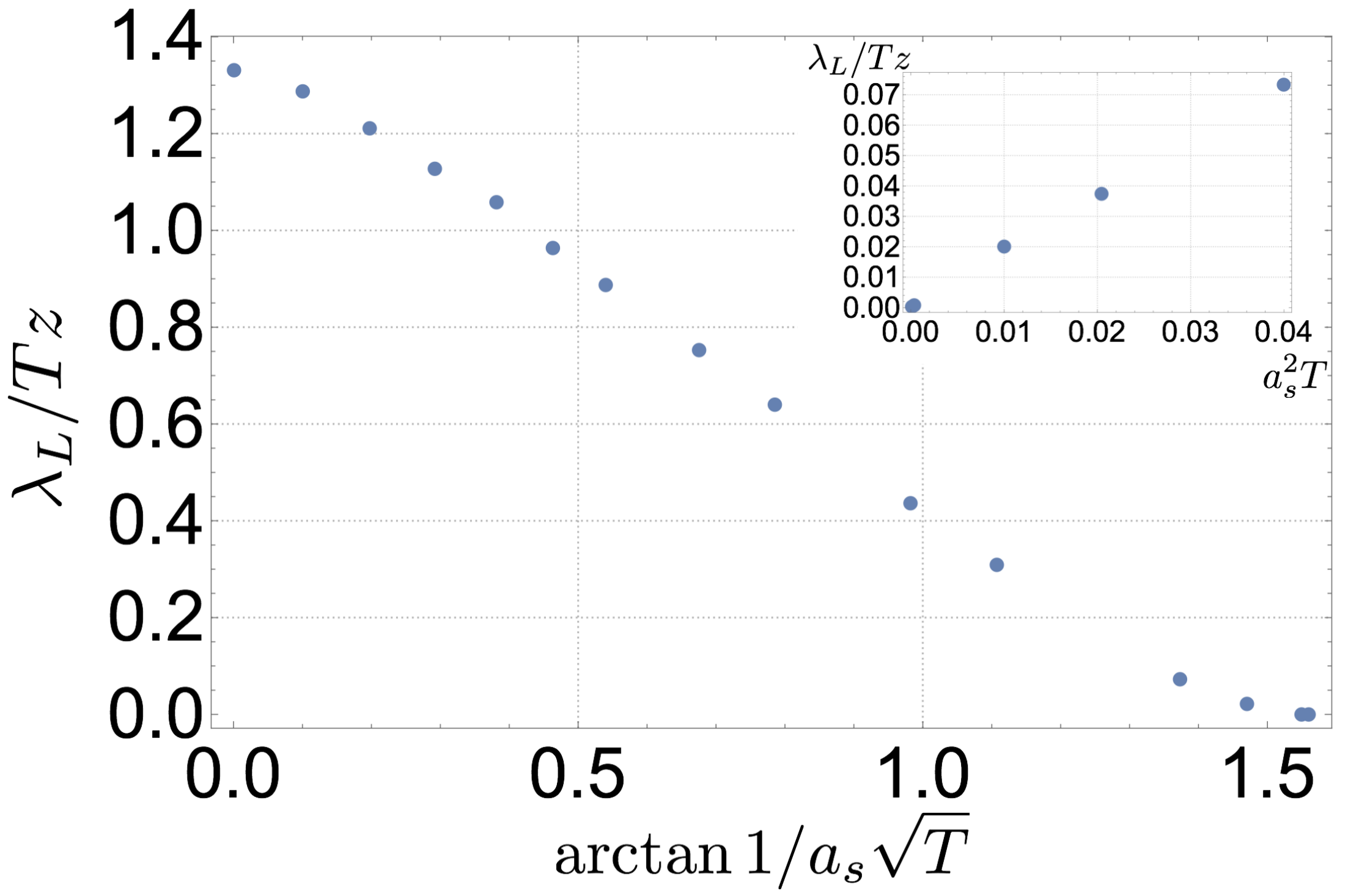}
 	\caption{Lyapunov exponents as a function of scattering length. This gives $\lambda_L \approx 1.32 z T$ in the unitary limit.}\label{fig:high}
 \end{figure}

To this leading order, the Lyapuonv exponent should be proportional to $z$. In the unitary limit with $a_s=\infty$, one expect $\lambda_L \propto z T$. We show simplified expressions directly used in numerics in Appendix. In the weakly interacting limit where $a_s\rightarrow 0^-$, $\lambda_L \propto z a_s^2 T^2$. Our results is symmetric for $a_s\rightarrow -a_s$ because we do not consider the distribution of bosons, which may be interpreted as considering physics in upper branch. Numerical results for Lyapunov exponents as a function of scattering length is shown in Fig. \ref{fig:high}, where we have determined $\lambda_L \approx 1.32 z T\approx 21\frac{n}{T^{1/2}}$ in the unitary limit, in which $n$ is the density for a single spin component. The result is parametrically smaller than the chaos bound. 

We are interested in the combination $v^2\tau_L$, which can be compared to diffusion constant. In present case the typical velocity is thermal velocity and we find $v_T^2\tau_L\sim 14 T^{3/2}/n$ with $v_T^2=3T/2$ to be the typical velocity of the system. It is interesting to compare this result with the energy diffusion constant. The heat conductivity $\kappa$ is calculated in \cite{thermal} by a variational method of Boltzmann equations in high temperature limit, and the result is found to be $\kappa=\frac{225}{128\sqrt{\pi}}T^{3/2}$. By using the Einstein's relation $D_E=\kappa/c_V$ with heat capacity $c_V$, we find $D_E\sim0.33T^{3/2}/n\ll v_T^2\tau_L$. We will give arguments to this after studying low temperature case.

\section{Quantum chaos in low temperature limit: effective field theory} \label{V}
The calculation with microscopic fermionic model is not controlled in low temperature. As a result we choose to use the effective description in terms of phonons for the unitary Fermi gas in low termperature limit. This is reasonable since when the system is deep in the superfluid phase and the only low energy excitation is phonon. Due to the (non-relativistic) conformal symmetry of the system, the effective theory for phonons can determined up to several coefficients \cite{Son} which are then determined using certain approximations \cite{epsilon,det1,det2,det3}.

To the leading order in gradient expansion, the effective action for phonon field $\phi$ is:
\begin{align}
\mathcal{L}_\phi=\frac{1}{2}(\partial_0\phi)^2-\frac{v_s^2}{2}(\nabla\phi)^2-g_3\left[(\partial_0\phi)^3-9v_s^2\partial_0\phi(\nabla\phi)^2\right],
\end{align}
with $v_s^2=\frac{2\mu}{3}$ and $g_3=\frac{\pi v_s^{\frac{3}{2}}\xi^{\frac{3}{4}}}{3^{\frac{1}{4}}8\mu^2}$, where $\mu=\xi_{u}T_F$ and $T_F$ is determined by density of fermions in non-interacting limit. We take the Bertsch parameter $\xi_u\sim 0.4$ here \cite{Uexp}. We will firstly set $v_s=1$ and finally add them back by dimensional analysis.

In this action, the dispersion of phonon is linear with $\epsilon_\mathbf{k}=v_s k$. However, whether a real process $\phi(\mathbf{p_1}) \rightarrow \phi(\mathbf{p_2})+\phi(\mathbf{p_3}) $ (and its inverse) can occur will depend on the next-to-leading order correction to linear dispersion. In \cite{epsilon} it is found that the correction is $\epsilon_\mathbf{k}=v_s k+u k^2$ with a positive $u$, as a result the splitting of a single phonon into two is allowed by conservation laws and at low temperature physics should be dominated by such process. Here we keep to $u^0$ and all particles move in the same direction.

\begin{figure}[t]
 	\center
 	\includegraphics[width=0.8\columnwidth]{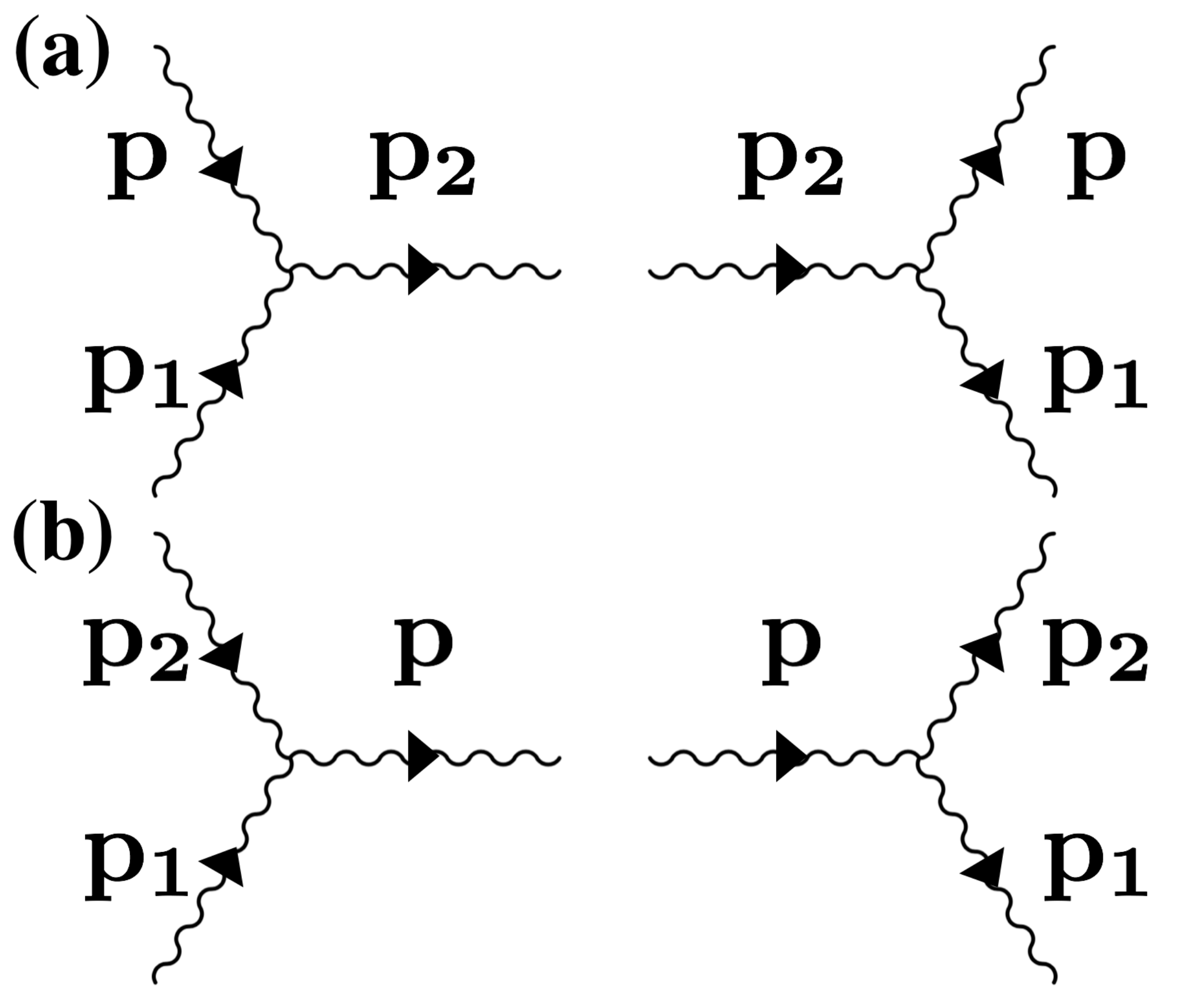}
 	\caption{Processes that govern the evolution of distribution function $F(\mathbf{p})$. Each row cancels out in thermal equilibrium.}\label{fig:Diagrams3}
\end{figure}

Now we would like to derive the generalized Boltzmann equation using our shortcut. Diagrams for the decay or formation of phonons are shown in Fig. \ref{fig:Diagrams3}.
Summing up these contributions, for homogeneous perturbation the evolution of distribution function $f(\mathbf{p},t)$ is described by \cite{thermal,Kamenev}:
\begin{align}
\frac{\partial f(\mathbf{p},t)}{\partial t}=I^{(a)}[\mathbf{p},f]+I^{(b)}[\mathbf{p},f], 
\end{align}
where the collision integrals are given by:
\begin{align}
I^{(b)}&=\int \frac{d\mathbf{p_1}d\mathbf{p_2}}{8 p p_1p_2(2\pi)^2}|\mathcal M(\mathbf{p},\mathbf{p_1};\mathbf{p_2})|^2\delta^{(4)}(p_2-p_1-p)\notag\\&[-f(\mathbf{p_1})(1+f(\mathbf{p_2}))f(\mathbf{p})+(1+f(\mathbf{p_1}))f(\mathbf{p_2})(1+f(\mathbf{p}))d],\\
I^{(b)}&=\frac{1}{2}\int \frac{d\mathbf{p_1}d\mathbf{p_2}}{8pp_1p_2(2\pi)^2}|\mathcal M(\mathbf{p_1},\mathbf{p_2};\mathbf{p})|^2\delta^{(4)}(p_2+p_1-p)\notag\\&[-(1+f(\mathbf{p_1}))(1+f(\mathbf{p_2}))f(\mathbf{p})+f(\mathbf{p_1})f(\mathbf{p_2})(1+f(\mathbf{p}))],
\end{align}
where the difference of factor is from the symmetry factor. By using the fact that the momentum of all particles is parallel, we have the scattering amplitude $|\mathcal M(\mathbf{p_1},\mathbf{p_2};\mathbf{p})|^2=(48 g_3 p p_1 p_2)^2$. Now we proceed to perform step 2-4 of the shortcut, and the result is:
\begin{align}
\frac{\partial F(\mathbf{p},t)_{mm'}}{\partial t}=\text{St}_{mm'}(\mathbf{p}),
\end{align}
with 
\begin{align}
\text{St}_{mm'}(\mathbf{p})=&\text{St}^{(1)}_{mm'}(\mathbf{p})+\text{St}^{(2)}_{mm'}(\mathbf{p})\\
\text{St}^{(1)}_{mm'}(\mathbf{p})=&\frac{1}{2}\int \frac{d\mathbf{p_1}d\mathbf{p_2}}{8pp_1p_2(2\pi)^2}|\mathcal M(\mathbf{p},\mathbf{p_1};\mathbf{p_2})|^2\delta^{(4)}(p_2-p_1-p)\notag \\&[(F(\mathbf{p_2})-F(\mathbf{p_1}))F_{mm'}(\mathbf{p})+F_{m'm}(\mathbf{p_1})F_{mm'}(\mathbf{p_2})], \label{boson1}\\
\text{St}^{(2)}_{mm'}(\mathbf{p})=&\frac{1}{4}\int \frac{d\mathbf{p_1}d\mathbf{p_2}}{8pp_1p_2(2\pi)^2}|\mathcal M(\mathbf{p_1},\mathbf{p_2};\mathbf{p})|^2\delta^{(4)}(p_2+p_1-p)\notag \\&[(-F(\mathbf{p_2})-F(\mathbf{p_1}))F_{mm'}(\mathbf{p})+F_{mm'}(\mathbf{p_1})F_{mm'}(\mathbf{p_2})].\label{boson2}
\end{align}
The label of $m $ and $m'$ is read out from the one-loop self energy diagram. For real boson, in thermal equilibrium we have:
\begin{align}
F^0(\mathbf{p})=1+2n_B(p),\ \ \ F^0_{ud}(\mathbf{p})=F^0_{du}(\mathbf{p})=\frac{1}{\sinh(p/2T)}.
\end{align}
One could check Eq. \eqref{boson1} and \ref{boson2} vanish for such solutions. The $\delta$ function is easily integrated out and we solve the eigenvalue of linearized generalized Boltzmann equations numerically. Now we assume $F_{ud}=F_{du}$ for all time. By power-counting and putting back $v_s$, the result should be $\lambda_L= C g_3^2T^5/T_F^4$. The numerical factor $C$ is found to be $C\approx 9\times10^3$. The Lyapunov exponent in low temperature then decreases much quicker ($\sim T^5$) than the chaotic bound ($\sim T$).

Because the typical velocity scale $v_s$ doesn't depend on temperature, we have $v_s^2\tau_L\sim 1/T^5$. However, $D_E=\kappa/c_V\sim 1/T$ because as found in \cite{thermal,abs}, $\kappa\sim T^2$ and for phonon gas $c_V \sim T^3$. Again, we find that $D_E\ll v_s^2\tau_L$ in low energy limit. 

We attribute such results to the momentum conservation which reduces the efficiency of energy transport significantly. For example, in the low energy limit, to the leading order the energy current by phonons is given by $J_E=\int v_p |p|f(\mathbf{p})$. However, if the dispersion is strictly linear $v_p=v_s$ then this vanishes due to the momentum conservation, if we start from an initial state with vanishing total momentum. Previous study of heat capacity due to phonons in low temperature limit of the Unitary Fermi gas indeed consider the correction of dispersion beyond linear \cite{thermal,abs}. Similarly, in the high temperature case $J_E=\int v_p \frac{p^2}{2}f(\mathbf{p})$. If we approximate the energy $p^2/2$ as $3T/2$, which is the expectation from thermal distribution, it again vanishes, indicating a large part of the thermal energy can not lead to thermal transport. Nevertheless this suppression is much larger for the phonon case, where we see a parametric difference between $D_E$ and $ v_s^2\tau_L$. We expect this to be a general mechanism for systems with quasiparticles.

\section{Summary and Outlooks}\label{VI}
In this work we have studied the chaotic behavior of the unitary Fermi gas in high and low temperature limit by using the generalized Boltzmann equations. In high temperature limit, we use the microscopic model and find $\lambda_L=21\frac{n}{T^{1/2}}$ in the unitary limit. In the low temperature limit, we utilize the effective field theory of phonons where $\lambda_L=9\times 10^3\left(\frac{T}{T_F}\right)^4T$. By comparing with previous results, we find $D_E\ll v^2\tau_L$ with typical velocity scale $v$ and we explain this as a result of the momentum conservation. We also propose a shortcut to the generalized Boltzmann equation by using the traditional Boltzmann equation.

One interesting question is whether we relate the calculation for quantum Lyapunov exponent from generalized Boltzmann equation to classical chaos, since in classical system we can also write down traditional Boltzmann equation and then modify them to write out the equation for chaos. This may provide more understanding about the relation between classical and quantum chaos. Another problem is to exploring the full temperature regime in certain large-N generalization. It is interesting if one could write our some matrix model which is related to the unitary Fermi gas where the Lyapunov exponent is not suppressed by $1/N$ factor.

 \textit{Acknowledgment.} We thank Yu Chen for dicussions. This research was supported in part by the National Science Foundation under Grant No. NSF PHY-1748958 and the Heising-Simons Foundation.

\appendix
\section{Relation between OTOC and generalized Boltzmann equation}\label{A}
We could get some intuition of using some kind of Boltzmann equations to study the behavior of OTOC. To avoid possible singularity, one could assume some explicit cut-off or take some lattice model. To begin with, we consider a doubled system prepared in a thermal field doubled state at inverse temperature $\beta$ \cite{TFD}:
\begin{align}
|\psi\rangle=\sum_n \exp(-\beta E_n/2)|n_u\rangle|n_d\rangle/\mathcal Z_\beta.
\end{align}
Here we use $u/d$ to label different Hilbert space. Now we perturb the system by applying an operator $W_u(0)$ in the $u$ system, then we have:
\begin{align}
|\tilde\psi(t=0) \rangle=W_u(0)|\psi\rangle.
\end{align}
Now we begin to evolve the system. Instead of using $H=H_u+H_d$, here we choose to use $H=H_u-H_d$, which makes $|\psi\rangle$ an eigenstates of the Hamiltonian. Then after time $t$ we do some measurements. We choose to measure operator $V_u^\dagger V_u$ or the correlation function $V_u^\dagger V_d$. For $V_u^\dagger V_u$, we have:
\begin{align}
\langle\tilde\psi(t=0)|V_u^\dagger(t) V_u(t)|\tilde\psi(t=0) \rangle=\left< W^\dagger(0) V^\dagger(t) V(t) W(0)\right>_\beta.\label{TFD1}
\end{align}
In the last equation, the measurement is done in a single system with temperature $\beta$. If we take the $V$ and $W$ to be the annihilation operator of particles, this is just an experiment of kicking out one atom and then study the evolution of particle density. In standard semi-classical approximation \cite{Kamenev}, the evolution of this density distribution can be described by the traditional Boltzmann equation, with an initial value determined by its value at $t=0$. This is the $uu$ part discussed in Eq. \eqref{ori}.

For the $u/d$ correlation function $V_u^\dagger V_d$, we have:
\begin{align}
\langle\tilde\psi(t=0)|V_u^\dagger(t)& V_d(t)|\tilde\psi(t=0) \rangle\notag\\&=\text{tr}\left[ \sqrt{\rho}W^\dagger(0) V^\dagger(t)W(0) \sqrt{\rho}V(t) \right]/\mathcal{Z}.\label{TFD2}
\end{align}
This turns out to be an OTOC \footnote{In fact here the imaginary time for the operator $V$ is different from Eq. \eqref{OTOC}. Nevertheless, we expect both definitions lead to similar behavior the same Lyapunov exponents in the long time limit. One way to see this is by realizing the homogeneous part of the self consistent equation for them in long time limit should be the same.}. The similarity between Eq.\eqref{TFD1} and \eqref{TFD2} suggest an unified semi-classical equation may exist, which can describe the evolution of OTOC from its initial condition at small $t$. This is the idea of generalized Boltzmann equation. The scrambling of information, which is described by the vanish of OTOC, implies the TFD, after the thermalization of this doubled system, will behavior like a tensor product of two thermal ensembles locally.

\section{simplified expressions for generlized Boltzmann equations in the high temperature limit}\label{B}
Here we give the simplified expressions used directly in numerics for generlized Boltzmann equations in the high temperature limit. We assume that for the mode with maximal Lyapunov exponent $F_{mm'}(\mathbf{p})=F_{mm'}(p)$ is rotational invariant and satisfy $F_{ud}=-F_{du}$ motivated by the unperturbed solution. In $\text{St}_{ud}[\mathbf{p},F]$, the term proportional to $\delta F_{ud}(p)$ is given by:
\begin{align}
\int_0^\infty\frac{2p_1^2dp_1}{\pi}f(p,p_1)(-4e^{\frac{\mu}{T}})e^{-\frac{p_1^2}{2T}}\delta F_{ud}(p),
\end{align}
with
\begin{align}
f(p,p_1)=&\frac{2}{a_s^4p p_1}[a_s^2(2|p-p_1|+a_s^2pp_1-2|p+p_1|)\notag\\&-8\log\frac{4+a_s^2|p-p_1|}{4+a_s^2|p+p_1|}].
\end{align}
The term proportional to $\delta F_{ud}(p_1)$ is given by:
\begin{align}
\int_0^\infty\frac{2p_1^2dp_1}{\pi}f(p,p_1)(4e^{\frac{\mu}{T}})e^{-\frac{p_1^2+p^2}{4T}}\delta F_{ud}(p_1),
\end{align}
where we have used $F_{ud}=-F_{du}$. Finally, for the last term proportional to $F_{ud}(p_2)$, we have:
\begin{widetext}
\begin{align}
\int\frac{t^2p_2^2 dp_2d\cos \theta_1d\cos \theta_2d\phi_1 dt}{(2\pi)^5}&\frac{16\pi^2a_s^2}{4+t^2a_s^2}8e^{\frac{\mu}{T}}e^{-\frac{2p^2-p_2^2+t^2+2 t p \cos\theta_1}{4T}} \delta F_{ud}(p_2)\times\notag\\&\delta[(p_2(\sin\theta_1\sin\theta_2\cos\phi_1+\cos\theta_1\cos\theta_2)-p \cos\theta_1)t-(p^2+p_2^2-2p p_2\cos\theta_2)].
\end{align}
\end{widetext}
where we could further integrate over $t$ by solving the constrain imposed by the $\delta$ function. In numerics, we only find a single positive eigenvalue and the Lyapunov exponent is well defined.

\end{document}